\documentclass[12pt]{article}

\input epsf
\usepackage{amssymb}
\usepackage[dvips]{graphicx}
\usepackage{amsmath,amscd}

\def\dj{\hbox{d\kern-0.347em \vrule width 0.3em height 1.252ex depth
-1.21ex \kern 0.051em}}

\numberwithin{equation}{section}

\begin{document}

\setlength{\oddsidemargin}{0cm}
\setlength{\baselineskip}{7mm}


\thispagestyle{empty}
\setcounter{page}{0}

\begin{flushright}
CERN-PH-TH/2008-224  \\
IFT-UAM/CSIC-08/80    \\
\end{flushright}

\vspace*{0.5cm}

\begin{center}
{\bf \Large Critical formation of trapped surfaces in the collision}

\vspace*{0.3cm}

{\bf \Large of gravitational shock waves}

\vspace*{0.7cm}

Luis \'Alvarez-Gaum\'e,$^{\,\rm a,}$\footnote{E-mail: 
{\tt Luis.Alvarez-Gaume@cern.ch}}
C\'esar G\'omez,$^{\rm\, b,}$\footnote{E-mail: 
{\tt Cesar.Gomez@uam.es}} 
Agust\'{\i}n Sabio Vera,$^{\rm\, a,}$\footnote{E-mail: 
{\tt Agustin.Sabio.Vera@cern.ch}} 
Alireza Tavanfar$^{\rm\, c,}$\footnote{E-mail: 
{\tt tavanfar@physics.harvard.edu}} 

and 
Miguel A. V\'azquez-Mozo$^{\rm\, d,}$\footnote{E-mail: 
{\tt Miguel.Vazquez-Mozo@cern.ch}}

\vspace*{0.25cm}

\begin{quote}

$^{\rm a}$\,\,{\sl Theory Group, Physics Department, CERN, CH-1211
Geneva 23, Switzerland}

$^{\rm b}$\,\,{\sl Instituto de F\'{\i}sica Te\'orica UAM/CSIC, 
Universidad Aut\'onoma de Madrid, E-28049 Madrid, Spain}

$^{\rm c}$\,\,{\sl Jefferson Physical Laboratory, Harvard University, Cambridge, MA 02138, 
U.S.A.}

$^{\rm d}$\,\, {\sl Departamento de F\'{\i}sica Fundamental, Universidad de 
Salamanca, Plaza de la Merced s/n, E-37008 Salamanca, Spain}

\end{quote}

\vspace*{0.4cm}
{\bf \large Abstract}
\end{center}

\noindent
We study the formation of marginally trapped surfaces in the head-on collision of two shock waves
both in anti-de Sitter and Minkowski space-time in various dimensions as a function of the spread of the
energy density in transverse space. For $D=4$ and $D=5$ it is shown that there exists a critical value of this spread
above which no solution to the trapped surface equation is found. In four dimensions the trapped surface
at criticality has a finite size, whereas in five the critical size vanishes. In both cases we find
scaling relations characterized by critical exponents. Finally, when $D>5$ there is always a solution to the
trapped surface equation for any transverse spread.

\vspace*{2cm}

\newpage

\setcounter{footnote}{0}

\section{Introduction}

Critical behavior in gravitational collapse has been widely studied in General Relativity 
since its discovery by Choptuik \cite{choptuikPRL} (see
\cite{reviews} for an updated review). Recently, 
the emergence and relevance of this phenomenon has been 
studied also in other realms. As an example, in \cite{us} a proposal was put forward for a holographic description of 
critical gravitational collapse
in terms of high energy scattering in gauge theories. In \cite{vw} Choptuik-like critical behavior was found in the context 
of transplanckian scattering \cite{acv}.

In this letter we present a detailed study of the emergence of critical behavior in the formation of a marginal trapped surface 
in the collision of two gravitational shock waves, both in anti-de Sitter (AdS) and Minkowski space-time. 
To summarize our results,  in AdS space we have found a dimension-dependent critical behavior in the formation of
a marginally trapped surface in the head-on collision of two shock waves with respect to the width of the waves in
transverse space. In both four and five dimensions we showed the existence of a critical value of this width 
above which the trapped surface is never formed. In the four-dimensional case the behavior found is what we could
denote as type I, i.e. the trapped surface presents a finite size at criticality and its scaling is characterized by a critical 
exponent $\gamma={1\over 2}$. In the five-dimensional case, on the other hand, 
we showed that the trapped surface has zero size at criticality. This type II scaling is described
by a critical exponent $\gamma=1$. Finally, for dimensions larger than five, we find that a marginally trapped surface is always formed for any value of the spread of the incoming waves.

In flat space-time the obtained results are qualitatively the same as those obtained for AdS. 
The only quantitative difference appears in 
five dimension where the critical exponent is $\gamma={1\over 2}$ instead of 1.

\section{Critical behavior in head-on shock wave collisions in AdS space-time}

Our aim in this section is to study the formation of closed trapped surfaces in the head-on collision
of two shock waves with an ``extension" in transverse space characterized by 
a parameter $\omega$. We start by studying the problem in AdS space, following the setup 
used in Ref. \cite{gpy}, to study afterwards the case without a cosmological constant.

Let us begin by briefly reviewing the results of Ref. \cite{gpy} with a slight change in notation.
We consider a shock wave propagating in $D$-dimensional AdS space-time, characterized by
the line element
\begin{eqnarray}
ds^{2}={L^{2}\over z^{2}}\Big(-dudv+d\vec{x}_{T}^{\,2}+dz^{2}\Big)+{L\over z}\Phi(z,\vec{x}_{T})\,
\delta(u)\,du^{2}.
\end{eqnarray} 
This metric is a solution to the Einstein equations for an energy momentum tensor with the single nonvanishing
component
\begin{eqnarray}
T_{uu}=\rho(z,\vec{x}_{T})\delta(u), 
\label{em_tensor}
\end{eqnarray}
provided the metric function $\Phi(z,\vec{x}_{T})$ satisfies the equation
\begin{eqnarray}
\left(\Box_{\mathbb{H}_{D-2}}-{D-2\over L^{2}}\right)\Phi(z,\vec{x}_{T})=
-16\pi G_{N}{z\over L}\rho(z,\vec{x}_{T}),
\label{ee1}
\end{eqnarray}
where $\Box_{\mathbb{H}_{D-2}}$ is the Laplace operator on the $(D-2)$-dimensional
hyperbolic space $\mathbb{H}_{D-2}$
\begin{eqnarray}
\Box_{\mathbb{H}_{D-2}}\equiv {z^{D-2}\over L^{2}}\partial_{z}\Big(z^{4-D}\partial_{z}\,\,\,\,\Big)
+{z^{2}\over L^{2}}\vec{\nabla}^{2}_{T}.
\end{eqnarray}

We are interested in solutions with O($D-2$) symmetry. This means that the function $\Phi(z,\vec{x}_{T})$
depends on the transverse coordinates only through the so-called ``chordal" coordinate $q$ defined 
as\footnote{Geometrically, this coordinate represents the square distance in embedding space, 
measured in units of $4L^{2}$, between the points $(z,\vec{x}_{T})$ and $(L,\vec{0})$ of $\mathbb{H}_{D-2}$.}
\begin{eqnarray}
q\equiv {(z-L)^{2}+\vec{x}_{T}^{\,2}\over 4Lz}.
\label{chordaldef}
\end{eqnarray}
In order to have solutions to (\ref{ee1}) with the required invariance we introduce the rescaled 
density
\begin{eqnarray}
\overline{\rho}(z,\vec{x}_{T})\equiv {z\over L}\rho(z,\vec{x}_{T}),
\label{rhobar}
\end{eqnarray}
which we take to depend only on $q$. Writing the line element of $\mathbb{H}_{D-2}$ in terms of the
chordal coordinate $q$
\begin{eqnarray}
ds^{2}_{\mathbb{H}_{D-2}}=L^{2}\left[{dq^{2}\over q(q+1)}+4q(q+1)d\Omega^{2}_{D-3}\right],
\end{eqnarray}
we find that Eq. (\ref{ee1}) can be written as
\begin{eqnarray}
q(q+1)\Phi''(q)+{1\over 2}(D-2)(1+2q)\Phi'(q)-(D-2)\Phi(q)=-16\pi G_{N}L^{2}\overline{\rho}(q).
\label{eqdiff1}
\end{eqnarray}

As shown in \cite{gpy} the solution to the previous equation behaves asymptotically for large $q$ as
\begin{eqnarray}
\Phi(q)&\sim &{8\pi G_{N}L^{2}\over D-1}\Phi_{+}(q)\int_{0}^{\infty}dq' (1+2q')[q'(1+q')]^{D-4\over 2}
\overline{\rho}(q')\nonumber \\
&=& {2^{6-D}L^{4-D}\pi G_{N}\over {\rm Vol}(S^{D-3})(D-1)}\,E\,\Phi_{+}(q),
\label{assympphi}
\end{eqnarray}
where 
\begin{eqnarray}
\Phi_{+}(q)=q^{2-D}{}_{2}F_{1}\left(D-2,{D\over 2};D;-{1\over q}\right)
\end{eqnarray}
and the energy $E$ is defined by
\begin{eqnarray}
E=2^{D-3}L^{D-2}{\rm Vol}(S^{D-3})\mu\int_{0}^{\infty}dq'(1+2q')[q'(1+q')]^{D-4\over 2}\overline{\rho}(q').
\label{energy}
\end{eqnarray}

The next question to be discussed is whether or not a black hole is formed as the result of 
the head-on collision of two waves of the type described above. In the region 
$\{u<0\}\cup \{v<0\}$, i.e. the part of the space-time previous to the collision, 
the metric is given by
\begin{eqnarray}
ds^{2}&=&{L^{2}\over z^{2}}\Big(-dudv+d\vec{x}_{T}^{\,2}+dz^{2}\Big)\nonumber \\
&+&{L\over z}\Phi_{1}(z,\vec{x}_{T})
\theta(v)\delta(u)du^{2}+{L\over z}\Phi_{2}(z,\vec{x}_{T})
\theta(u)\delta(v)dv^{2}.
\label{lineelementbc}
\end{eqnarray} 
A rigorous analysis of the formation of a black hole in the collision would require to solve
the Einstein equations in the interaction region $\{u>0,v>0\}$ (see, for example, 
\cite{GR}). A sufficient condition for black hole formation, however, is the existence of 
a marginal closed trapped surface in the hypersurface $\{u\leq 0,v=0\}\cup \{u=0,v\leq 0\}$
\cite{penrose,dp,eardley_giddings,kv,gpy}. In order to find this trapped surfce,
however, the coordinates used to write the
line element (\ref{lineelementbc}) are not very convenient since null geodesics are discontinuous
across the wave fronts $u=0$, $v=0$. This can be avoided by switching to a new system of 
coordinates $(U,V,Z,\vec{X}_{T})$ in which the delta function terms are eliminated in the metric
(see \cite{dp,GR}) and the geodesics are continuous. 

In this new system of coordinates the trapped surface we are looking for has two parts that we 
denote by $\mathcal{S}_{1}$ and $\mathcal{S}_{2}$ and respectively lie in the regions $V<0$ and 
$U<0$. They are defined in terms of 
two functions $\psi_{1}(Z,\vec{X}_{T})$ and $\psi_{2}(Z,\vec{X}_{T})$ by
\begin{eqnarray}
\mathcal{S}_{1}:\left\{
\begin{array}{l}
U=0 \\
V+\psi_{1}(Z,\vec{X}_{T})=0
\end{array}
\right., \hspace*{1cm}
\mathcal{S}_{2}:\left\{
\begin{array}{l}
V=0 \\
U+\psi_{2}(Z,\vec{X}_{T})=0
\end{array}
\right.,
\label{ctp}
\end{eqnarray}
with the additional boundary conditions at the intersection $\mathcal{C}=\{U=V=0\}$
\begin{eqnarray}
\psi_{1}(Z,\vec{X}_{T})\Bigg|_{\mathcal{C}}=0, \hspace*{1cm}
\psi_{2}(Z,\vec{X}_{T})\Bigg|_{\mathcal{C}}=0
\label{bc1}
\end{eqnarray}
and
\begin{eqnarray}
\Big[\partial_{Z}\psi_{1}\partial_{Z}\psi_{2}+(\vec{\nabla}\psi_{1})\cdot(\vec{\nabla}\psi_{2})\Big]
\Bigg|_{\mathcal{C}}=4.
\label{bc2}
\end{eqnarray}
Since $\mathcal{S}_{1}$ and $\mathcal{S}_{2}$ lie respectively in the regions $V<0$ and $U<0$, we also have
that $\psi_{1}(Z,\vec{X}_{T})>0$, $\psi_{2}(Z,\vec{X}_{T})>0$. 

The two functions $\psi_{1}(Z,\vec{X}_{T})$ and $\psi_{2}(Z,\vec{X}_{T})$ 
have to be determined by imposing the condition that the surface they
define is marginally trapped \cite{poisson}, i.e. that the congruence of outgoing null 
geodesics orthogonal to the surface has zero expansion. We consider a symmetric head-on
collision where $\Phi_{1}(Z,\vec{X}_{T})=\Phi_{2}(Z,\vec{X}_{T})\equiv \Phi(Z,\vec{X}_{T})$
and $\psi_{1}(Z,\vec{X}_{T})=\psi_{2}(Z,\vec{X}_{T})\equiv \psi(Z,\vec{X}_{T})$
and define the rescaled function
\begin{eqnarray}
\Psi(Z,\vec{X}_{T})\equiv {L\over Z} \psi(Z,\vec{X}_{T}).
\end{eqnarray}
In the O($D-2$)-symmetric case this function depends only on the chordal coordinate\footnote{On the 
hypersurfaces $\{U=0,V<0\}$ and $\{V=0,U<0\}$ we have $Z=z$ and $\vec{X}_{T}=\vec{x}_{T}$.
Then the chordal coordinate $q$ in this case is also given by Eq. (\ref{chordaldef}).}.
Then the condition of zero expansion reduces to the equation \cite{gpy}
\begin{eqnarray}
\left[q(1+q)\partial_{q}^{2}+{1\over 2}(D-2)(1+2q)\partial_{q}-(D-2)
\right]\Big[\Phi(q)-\Psi(q)\Big]=0,
\label{eqpsi}
\end{eqnarray}
supplemented by the boundary conditions (\ref{bc1}) and (\ref{bc2}) which
now are the following boundary conditions at $q=q_{\mathcal{C}}$
\begin{eqnarray}
\Psi(q_{\mathcal{C}}) =0, \hspace{1cm} \Psi'(q_{\mathcal{C}})=-{2L\over \sqrt{q_{\mathcal{C}}
(1+q_{\mathcal{C}})}}.
\label{bcq}
\end{eqnarray}
The function $\Phi(q)$ appearing in equation (\ref{eqpsi}) is known and given in terms of the
density $\overline{\rho}(q)$.
Hence, Eq. (\ref{eqpsi}) can be solved subjected to the boundary conditions (\ref{bcq}). A solution to this problem,
and therefore a marginally closed trapped surface, exists whenever there is a value of 
$q_{\mathcal{C}}$ solving the transcendental equation \cite{gpy}
\begin{eqnarray}
8\pi G_{N}L\int_{0}^{{q}_{\mathcal{C}}}dq[q(1+q)]^{D-4\over 2}(1+2q)
\overline{\rho}(q)=(1+2q_{\mathcal{C}})[q_{\mathcal{C}}(1+q_{\mathcal{C}})]^{D-3\over 2}.
\label{eqbasic1}
\end{eqnarray}

In Ref. \cite{gpy} the situation considered was the head-on collision of two identical shock waves corresponding
to the gravitational field created by a massless point particle with energy $\mu$, 
\begin{eqnarray}
\overline{\rho}(q)={2(2L)^{2-D}\over {\rm Vol}(S^{D-3})}{\mu\over [q(1+q)]^{D-4\over 2}}
\delta(q-\epsilon).
\label{rhobardelta}
\end{eqnarray}
where $\epsilon\rightarrow 0^{+}$ shifts the support of the delta function into the $q>0$ domain.
From Eq. (\ref{eqbasic1}), 
the existence of the marginally closed trapped surface is determined by the solution to the
equation
\begin{eqnarray}
\Big[q_{\mathcal{C}}(1+q_{\mathcal{C}})\Big]^{D-3\over 2}(1+2q_{\mathcal{C}})={8\pi G_{N}\mu
(2L)^{3-D}\over {\rm Vol}(S^{D-3})}.
\end{eqnarray}
The function on the left-hand side of this equation grows monotonically from zero and asymptotically 
behaves as $q_{\mathcal{C}}^{D-2}$. Hence, there always exists a (unique) solution to the equation and
therefore a marginally trapped surface.

In what follows, however, we would like to consider the situation in which the colliding shock waves 
describe the infinite boost of an extended source. This corresponds to a function ${\rho}(z,\vec{x}_{T})$ 
whose support does not have zero measure in the hyperbolic space $\mathbb{H}_{D-2}$ \cite{gpy}.
To preserve O($D-2$) symmetry, we take the following form for the 
rescaled density $\overline{\rho}(q)$
\begin{eqnarray}
\overline{\rho}(q)={2(2L)^{2-D}\over {\rm Vol}(S^{D-3})}{\mu\over [q(1+q)]^{D-4\over 2}}F(\omega,q),
\label{smearingrho}
\end{eqnarray}
where $\omega>0$ is a parameter and the function $F(\omega,q)>0$ is integrable and normalized according to
\begin{eqnarray}
\int_{0}^{\infty}dq\, F(\omega,q)=1.
\label{integrab}
\end{eqnarray}
In addition, we require the wave profile $\Phi(q)$ to have a well behaved asymptotic limit as
$q\rightarrow\infty$ or, in other words, to have finite energy. Then, from Eq. (\ref{energy}) we derive the condition
\begin{eqnarray}
\int_{0}^{\infty}dq\, (1+2q)F(\omega,q)<\infty.
\label{finiteenergy}
\end{eqnarray}
In the following we are going to focus on functions $F(\omega,q)$
which regularize the delta function $\delta(q-\epsilon)$ in Eq. (\ref{rhobardelta}), i.e.
\begin{eqnarray}
\lim_{\omega\rightarrow 0^{+}}F(\omega,q)=\delta(q-\epsilon)
\label{deltaregcond}
\end{eqnarray}
and that satisfy the condition (\ref{finiteenergy}). The parameter $\omega$, which measures the 
width of the energy distribution in transverse space, can be interpreted here as a kind of diluting parameter.

Using (\ref{smearingrho}) the condition for the marginally trapped surface now reads
\begin{eqnarray}
{8\pi G_{N}\mu (2L)^{3-D}\over {\rm Vol}(S^{D-3})}\int_{0}^{q_{\mathcal{C}}}dq\,
(1+2q)F(\omega,q)=(1+2q_{\mathcal{C}})[q_{\mathcal{C}}(1+q_{\mathcal{C}})]^{D-3\over 2}.
\end{eqnarray}
It proves to be convenient to make a change of variables in the integral to eliminate the dependence of the
integration limits on $q_{\mathcal{C}}$. This gives
\begin{eqnarray}
q_{\mathcal{C}}\int_{0}^{1}du\,(1+2q_{\mathcal{C}}u)F(\omega,q_{\mathcal{C}}u)= {(2L)^{D-3}{\rm Vol}(S^{D-3})
\over 8\pi G_{N}\mu} 
(1+2q_{\mathcal{C}})[q_{\mathcal{C}}(1+q_{\mathcal{C}})]^{D-3\over 2}.
\label{keyeqAdS}
\end{eqnarray}
Written in this way the left-hand side of the equation is independent of the dimension and its
global behavior can be easily studied.  First of all, from the finite energy condition (\ref{finiteenergy}) we have
\begin{eqnarray}
\lim_{q_{\mathcal{C}}\rightarrow \infty}
q_{\mathcal{C}}\int_{0}^{1}du\,(1+2q_{\mathcal{C}}u)F(\omega,q_{\mathcal{C}}u)
=\lim_{q_{\mathcal{C}}\rightarrow\infty}
\int_{0}^{q_{\mathcal{C}}}dq\,(1+2q)F(\omega,q)=\mbox{constant},
\end{eqnarray}
i.e., the left-hand side of Eq. (\ref{keyeqAdS}) saturates to a constant for large values of $q_{\mathcal{C}}$. 
On the other hand, for small $q_{\mathcal{C}}$ we have
\begin{eqnarray}
q_{\mathcal{C}}\int_{0}^{1}du\,(1+2q_{\mathcal{C}}u)F(\omega,q_{\mathcal{C}}u)\sim
F(\omega,0)q_{\mathcal{C}}.
\label{smallqint}
\end{eqnarray}
Then we find that 
the left-hand side of Eq. (\ref{keyeqAdS}) is a monotonous growing function of $q_{\mathcal{C}}$ that
behaves linearly at $q_{\mathcal{C}}\rightarrow 0^{+}$ and saturates to an $\omega$-dependent 
constant for $q_{\mathcal{C}}
\rightarrow \infty$. The discussion of the existence of solutions to Eq. (\ref{keyeqAdS}) 
has to be done separately for different
dimensions.

\paragraph{$\boldsymbol{D=4}$.} In four dimensions Eq. (\ref{keyeqAdS}) reads
\begin{eqnarray}
q_{\mathcal{C}}\int_{0}^{1}du\,(1+2q_{\mathcal{C}}u)F(\omega,q_{\mathcal{C}}u)= {{L} 
\over 2 G_{N}\mu} (1+2q_{\mathcal{C}})\sqrt{q_{\mathcal{C}}(1+q_{\mathcal{C}})}.
\label{d4}
\end{eqnarray}
The right-hand side of this equation is a monotonous function of $q_{\mathcal{C}}$ that stars at
the origin with infinite slope and behaves asymptotically as $q_{\mathcal{C}}^{2}$. Depending on the value of
$\omega$, the curves defined by both sides of (\ref{d4}) have two, one or no intersecting points, apart from
the trivial solution $q_{\mathcal{C}}=0$.  We see that there is 
a critical value $\omega^{*}$ such that for $\omega>\omega^{*}$ there are no nontrivial solutions to
Eq. (\ref{d4}), whereas for $\omega<\omega^{*}$ the two curves meet at two finite values of $q_{\mathcal{C}}$. 
The critical value of $\omega$ is determined by the condition that the two curves osculate at a critical value 
$q_{\mathcal{C}}^{*}$. This means that at $q_{\mathcal{C}}=q_{\mathcal{C}}^{*}$ both sides of Eq. (\ref{d4})  and their
first derivatives coincide. 

It is important to realize that the two solutions to Eq. (\ref{d4}) correspond to two trapped surfaces with different size 
$q_{\mathcal{C}}$. This means that one of the surfaces is inside the other. 
Since the apparent horizon is defined as the boundary of the trapped region we consider only the outermost 
marginally trapped surface, which corresponds to the upper branch of the solution.

To keep the analysis general let us
denote by $G(\omega,q_{\mathcal{C}})$ the difference between the two sides of Eq. (\ref{d4}). Then $\omega^{*}$
and $q_{\mathcal{C}}^{*}$ are determined by
\begin{eqnarray}
G(\omega^{*},q_{\mathcal{C}}^{*})=0, \hspace*{1cm} G^{(0,1)}(\omega^{*},q_{\mathcal{C}}^{*})=0,
\label{osculate}
\end{eqnarray}
where we have used an obvious notation for the partial derivatives. Since both sides of Eq. (\ref{d4}) are analytic
for $q_{\mathcal{C}}>0$, we can expand $G(\omega,q_{\mathcal{C}})$ around $\omega=\omega^{*}$ and 
$q_{\mathcal{C}}=q_{\mathcal{C}}^{*}$. Solving now the equation $G(\omega,q)=0$ and using then the conditions 
(\ref{osculate}) we find at leading order for $\omega\lesssim \omega^{*}$
\begin{eqnarray}
q_{\mathcal{C}}-q_{\mathcal{C}}^{*}\sim \sqrt{2 G^{(1,0)}(\omega^{*},q_{\mathcal{C}}^{*})\over 
G^{(0,2)}(\omega^{*},q_{\mathcal{C}}^{*})}(\omega^{*}-
\omega)^{1\over 2},
\label{cbd4}
\end{eqnarray}
where the argument of the square root is positive whenever a solution to the equations (\ref{osculate})
exists. 

To summarize, we have found that for $D=4$ there is a threshold in $\omega$ for the formation of the
marginally trapped surface defined by Eq. (\ref{ctp}). This surface only exists for values of 
$\omega$ smaller than the critical value $\omega^{*}$ and its size is finite at threshold. This is reminiscent
of the situation encountered in the so-called type I critical black hole formation \cite{reviews}. From Eq. 
(\ref{cbd4}) we find that in our case we find a critical exponent $\gamma={1\over 2}$. 

\paragraph{$\boldsymbol{D=5}$.}The equation determining
$q_{\mathcal{C}}$ is now
\begin{eqnarray}
q_{\mathcal{C}}\int_{0}^{1}du\,(1+2q_{\mathcal{C}}u)F(\omega,q_{\mathcal{C}}u)= {2L^{2}\over G_{N}\mu}
q_{\mathcal{C}}(1+2q_{\mathcal{C}})(1+q_{\mathcal{C}}).
\label{d5}
\end{eqnarray}
The function on the right-hand side grows monotonically and 
behaves linearly at $q_{\mathcal{C}}=0$ with slope ${2L^{2}\over G_{N}\mu}$
while at large $q_{\mathcal{C}}$ grows like $q_{\mathcal{C}}^{3}$. Therefore from Eq. (\ref{smallqint}) 
we find that the two curves on both sides of (\ref{d5})
intersect at a single point at finite $q_{\mathcal{C}}$ whenever
\begin{eqnarray}
{2L^{2}\over G_{N}\mu}<F(\omega,0).
\end{eqnarray}
For any regularization of the delta function, $F(\omega,0)$ monotonously decreases with $\omega$. 
Then, the previous equation determines a critical value of the diluting parameter $\omega^{*}$
\begin{eqnarray}
F(\omega^{*},0)={2L^{2}\over G_{N}\mu},
\end{eqnarray}
and a solution to Eq. (\ref{d5}) exists only when $\omega<\omega^{*}$, whereas for 
$\omega>\omega^{*}$ there is no marginal trapped surfaces of the type we are studying.

From Eq. (\ref{d5})  the size of the trapped surfaces vanishes for $\omega=\omega^{*}$, 
that is, $q_{\mathcal{C}}=0$. Hence, the scaling of $q_{\mathcal{C}}$ with $\omega^{*}-\omega$
for slightly subcritical  values of $\omega$ can be found by expanding the function $F(\omega,q_{\mathcal{C}})$
in  (\ref{d5}) in series about $q_{\mathcal{C}}=0$. Taking into account that for regularizations of 
the delta function we have that $F^{(0,1)}(\omega,0)=0$ we have
\begin{eqnarray}
F(\omega,q_{\mathcal{C}})={2L^{2}\over G_{N}\mu}-F^{(1,0)}(\omega^{*},0)(\omega^{*}-\omega)
+{1\over 2}F^{(0,2)}(\omega^{*},0)q_{\mathcal{C}}^{2}+\ldots
\end{eqnarray}
Plugging this expansion into Eq. (\ref{d5}) and solving for $q_{\mathcal{C}}$ we find that the scaling for 
$\omega\lesssim \omega^{*}$ 
is given by
\begin{eqnarray}
q_{\mathcal{C}}\sim {G_{N}\mu\over 4L^{2}} |F^{(1,0)}(\omega^{*},0)|(\omega^{*}-\omega)
\label{critd=5}
\end{eqnarray}

Our analysis of the five-dimensional case shows the existence of a critical value of the diluting parameter
$\omega$ above which there are no solutions to the equation for the marginally trapped surface. We find that
at $\omega^{*}$ there is a trapped surface of finite size, whereas for subcritical values
of $\omega$ we have a scaling characterized by a critical exponent $\gamma=1$. 
This behavior of the size of the marginally trapped surfaced formed in the collision is similar to the type II criticality 
found by Choptuik in black hole formation \cite{choptuikPRL,reviews}.

\paragraph{$\boldsymbol{D\geq 6}$.} From Eq. (\ref{keyeqAdS})
we see that the right-hand side behaves at the origin as $q^{D-3\over 2}_{\mathcal{C}}$,
where now the exponent is larger than one.
This means that the slope of the curve is zero at the origin. At the same time, for large $q_{\mathcal{C}}$ the function
grows as $q_{\mathcal{C}}^{D-2}$. Recalling the global behavior found above for the left-hand side of (\ref{keyeqAdS}) we
conclude that the two curves always meet at a single point and, as a consequence, a marginally
closed trapped surface of the sought type is always formed. In this case we find no critical behavior of any kind.

\section{The case of Minkowski space-time}

It is very interesting to know how the previous result
depend on the presence of a nonvanishing cosmological constant. To clarify this point, we study the head-on collision of two shock waves
in Minkowski space-time. The metric corresponding to the first metric is ($v<0$)
\begin{eqnarray}
ds^{2}=-dudv+d\vec{x}_{\perp}^{\,2}+\Phi_{1}(\vec{x}_{\perp})\delta(u)du^{2},
\end{eqnarray}
whereas the line element for the second wave ($u<0$) can be obtained from the previous equation 
by replacing $u\rightarrow v$ and $\Phi_{1}(\vec{x}_{\perp})\rightarrow \Phi_{2}(\vec{x}_{\perp})$. 

The whole analysis performed in the previous section can be carried
out also in Minkowski space-time. Equivalently, the equations for the O($D-2$)-symmetric case in flat space-time
can be obtained from the ones for AdS by noticing that
AdS space-time is locally equivalent to Minkowski space-time. We follow this approach.

Let us consider a small neighborhood of the point $z=L$, $\vec{x}_{T}=\vec{0}$ in $\mathbb{H}_{D-2}$. Defining $\overline{z}\equiv
z-L$ and studying the limit $|\vec{x}_{T}|\ll  L$, $\overline{z}\ll L$, we find that the chordal coordinate (\ref{chordaldef}) 
can be written as
\begin{eqnarray}
q={\overline{z}^{\,2}+\vec{x}_{T}^{\,2}\over 4L(\bar{z}+L)}\simeq {1\over 4L^{2}}\Big(\bar{z}^{\,2}+\vec{x}_{T}^{\,2}\Big)
\equiv {\vec{x}_{\perp}^{\,2}\over 4L^{2}},
\label{limiteq}
\end{eqnarray}
where we have defined the Minkowskian transverse coordinate $x_{\perp}^{a}=(\bar{z},x_{T}^{i})$, with $a=1,\ldots,D-2$. To keep 
the analogy with the AdS case, we introduce a fiducial length $\ell$ and define the dimensionless quantity
\begin{eqnarray}
p\equiv {\vec{x}_{\perp}^{\,2}\over 4\ell^{2}}.
\end{eqnarray}
We should stress that the scale $\ell$ is completely arbitrary.  
In the limit (\ref{limiteq}) the chordal coordinate is replaced by
\begin{eqnarray}
q \longrightarrow {\ell^{2}\over L^{2}}p.
\label{replacement}
\end{eqnarray}
Now, the Minkowski field equations can be obtained from Eq. (\ref{eqdiff1}) by making the replacement (\ref{replacement}) together
with $\Phi(q)\rightarrow \Phi(p)$ and taking the limit $L\rightarrow\infty$ at the end. Doing so we find\footnote{Unlike
in the case of the AdS equation (\ref{eqdiff1}), here there is no need to introduce a rescaled density. This can be seen 
by noticing that ${z\over L}=1+{\bar{z}\over L}$ tends to one in the limit $L\rightarrow\infty$ with $\bar{z}$ fixed.}
\begin{eqnarray}
p\Phi''(p)+{D-2\over 2}\Phi'(p)=-16\pi G_{N}\ell^{2}\rho(p).
\label{minkdiffeq}
\end{eqnarray}
The regular solution to the inhomogeneous problem (\ref{minkdiffeq}) can be written as
\begin{eqnarray}
\Phi(p)=\int_{0}^{\infty}dp'\,p'{}^{D-4\over 4}G(p,p')\rho(p').
\end{eqnarray}
with 
\begin{eqnarray}
G(p,p')={8\pi G_{N}\ell^{2}\over D-1}\Big[p'{}^{-{D-4\over 2}}\theta(p'-p)+p^{-{D-4\over 2}}\theta(p-p')\Big].
\end{eqnarray}
If the density $\rho(p)$ has compact support, or decreases fast enough at large $p$, the solution $\Phi(p)$
behaves asymptotically as
\begin{eqnarray}
\Phi(p)\sim {2^{6-D}\ell^{4-D}\pi G_{N}\over {\rm Vol}(S^{D-3})(D-1)}\,E\,p^{-{D-4\over 2}}, \hspace*{1cm}
p\rightarrow\infty,
\end{eqnarray}
where the energy $E$ is given by
\begin{eqnarray}
E=2^{D-3}\ell^{D-2}{\rm Vol}(S^{D-3})\int_{0}^{\infty}dp'\,p'{}^{D-4\over 2}\rho(p').
\label{energy_mink}
\end{eqnarray}

The calculation of marginally trapped surfaces in the collision of two shock waves in flat space-time has been done 
\cite{penrose,dp,eardley_giddings} and can be carried out following essentially the same steps taken above for the AdS case.
Changing coordinates from $(u,v,\vec{x}_{\perp})$ to $(U,V,\vec{X}_{\perp})$ \cite{dp} to eliminate the Dirac delta terms in the metric, 
we consider a candidate for the marginally trapped surface as the union of the two surfaces $\mathcal{S}_{1}$ and $\mathcal{S}_{2}$
defined by
\begin{eqnarray}
\mathcal{S}_{1}:\left\{
\begin{array}{l}
U=0 \\
V+\Psi_{1}(\vec{X}_{\perp})=0
\end{array}
\right., \hspace*{1cm}
\mathcal{S}_{2}:\left\{
\begin{array}{l}
V=0 \\
U+\Psi_{2}(\vec{X}_{\perp})=0
\end{array}
\right.,
\end{eqnarray}
where $\mathcal{S}_{1}$ lies in the region $V<0$ and $\mathcal{S}_{2}$ in $U<0$. 
The functions $\Psi_{1}(\vec{X})$ and $\Psi_{2}(\vec{X})$ are positive semidefinite and vanish only at the intersection of the
two surfaces, $\mathcal{C}=\{U=V=0\}$. We are interested in studying the case of the head-on collision of two identical waves,
$\Phi_{1}(\vec{X})=\Phi_{2}(\vec{X})$, where we have that $\Psi_{1}(\vec{X}_{\perp})
=\Psi_{2}(\vec{X}_{\perp})\equiv \Psi(\vec{X}_{\perp})$. Here we consider the O($D-2$)-symmetric case in 
which $\Phi(\vec{X}_{\perp})$ and $\Psi(\vec{X}_{\perp})$ depend only on $p={\vec{X}_{\perp}^{2}\over 4\ell^{4}}$. 
Then it can be shown that the surface $\mathcal{S}_{1}\cup
\mathcal{S}_{2}$ is a marginally trapped surface provided
\begin{eqnarray}
8\pi G_{N}\ell \int_{0}^{p_{\mathcal{C}}}dp'\,p'{}^{D-4\over 2}\rho(p')=p_{\mathcal{C}}^{D-3\over 2}
\label{tscm}
\end{eqnarray}
admits a solution for $p_{\mathcal{C}}$. This solution gives the radius of the surface at $\mathcal{C}$, the function 
$\Psi(p)$ determining $\mathcal{S}_{1}\cup \mathcal{S}_{2}$ being given by
\begin{eqnarray}
\Psi(p)=\Phi(p)-\Phi(p_{\mathcal{C}}).
\end{eqnarray}

As in the case of AdS, here we are going to study what happens when the density $\rho(p)$ is smeared in transverse space
with a width $\omega$. In technical terms, we consider a density of the form
\begin{eqnarray}
{\rho}(p)={2(2\ell)^{2-D}\over {\rm Vol}(S^{D-3})}{\mu\over p^{D-4\over 2}}F(\omega,p),
\label{smearingrhoMink}
\end{eqnarray}
where $F(\omega,p)$ is a smearing of the Dirac delta function satisfying Eqs. (\ref{integrab}) and (\ref{deltaregcond}). 
From the definition of the energy (\ref{energy_mink}) we see that the latter condition implies that the
corresponding solution has finite energy. The trapped surface condition (\ref{tscm}) then reads
\begin{eqnarray}
p_{\mathcal{C}}\int_{0}^{1}du \,F(\omega,p_{\mathcal{C}}u)={(2\ell)^{D-3}{\rm Vol}(S^{D-3})
\over 8 \pi G_{N}\mu}\, p_{\mathcal{C}}^{D-3\over 2}.
\label{geneqM}
\end{eqnarray}
Again, the left-hand side is independent of the dimension, their properties being very similar to the ones derived for the 
corresponding integral in the AdS case: it is a monotonously growing function of 
$q_{\mathcal{C}}$, with slope $F(\omega,0)$
at the origin and saturates to a constant for $q_{\mathcal{C}}\rightarrow\infty$. To find the solutions to the equation we
analyze the case for different dimensions.

\paragraph{$\boldsymbol{D=4}$.} In this case the equation to solve is
\begin{eqnarray}
p_{\mathcal{C}}\int_{0}^{1}du \,F(\omega,p_{\mathcal{C}}u)=
{\ell \over 2G_{N}\mu}\sqrt{p_{\mathcal{C}}}.
\label{d4m}
\end{eqnarray}
The situation is the same we found in the case of AdS$_{4}$. Namely, there is a critical value $\omega=\omega^{*}$
at which the curves in the left- and right-hand side of Eq. (\ref{d4m}) osculate. For $\omega>\omega^{*}$ the two 
curves do not cross at nonzero $\omega$ and as a consequence there is no solution of the trapped surface conditions.
For subcritical values of $\omega$ the curves meet at two points corresponding to two different values of $q_{\mathcal{C}}$. 
For the same reasons explained in the previous section, we choose the largest value.

The analysis of the behavior of the solutions when $\omega\lesssim \omega^{*}$ carried
out for AdS is completely general and can be applied to this case as well. 
It is found that the trapped surface has a nonvanishing size at criticality $p^{*}_{\mathcal{C}}$ with a
scaling characterized by a critical exponent $\gamma={1\over 2}$, as given in Eq. (\ref{critd=5}).

\paragraph{$\boldsymbol{D=5}$.} The structure of solutions of
\begin{eqnarray}
p_{\mathcal{C}}\int_{0}^{1}du \,F(\omega,p_{\mathcal{C}}u)={2\ell^{2}\over G_{N}\mu}q_{\mathcal{C}}
\label{d5m}
\end{eqnarray}
is analogous to the AdS case. This equation has a single solution whenever
\begin{eqnarray}
{2\ell^{2}\over G_{N}\mu}<F(\omega,0),
\end{eqnarray}
which implies the existence of a critical value $\omega^{*}$ such that for $\omega>\omega^{*}$ no solution to Eq. (\ref{d5m})
exists. To find the critical behavior for $\omega^{*}$ we notice that for $\omega=\omega^{*}$ the size of the 
trapped surface is zero and therefore Eq. (\ref{d5m}) can be solved for $\omega\lesssim\omega^{*}$ by expanding
$F(\omega,p_{\mathcal{C}})$ around $(\omega^{*},0)$. Plugging 
\begin{eqnarray}
F(\omega,p_{\mathcal{C}})={2\ell^{2}\over G_{N}\mu}-F^{(1,0)}(\omega^{*},0)(\omega^{*}-
\omega)+{1\over 2}F^{(0,2)}(\omega^{*},0)p_{\mathcal{C}}^{2}+\ldots
\label{expansionM}
\end{eqnarray}
into Eq. (\ref{d5m}) and solving the equation for $p_{\mathcal{C}}$ results in
\begin{eqnarray}
q_{\mathcal{C}}\sim  \sqrt{4F^{(1,0)}(\omega^{*},0)\over F^{(0,2)}(\omega^{*},0)}
(\omega^{*}-\omega)^{1\over 2}.
\label{d5M}
\end{eqnarray}
This means that the critical formation of the trapped surface is characterized by 
a critical exponent $\gamma={1\over 2}$. It is important to keep in mind that in writing
the expansion (\ref{expansionM}) we have used that $F(\omega,p_{\mathcal{C}})$ is 
an even function in $p_{\mathcal{C}}$. In addition, the square root in (\ref{d5M}) 
is always real since for any delta function regularization both $F^{(1,0)}(\omega^{*},0)$ and 
$F^{(0,2)}(\omega^{*},0)$ are negative.

\paragraph{$\boldsymbol{D\geq 6}$.} The situation in dimensions larger than five is 
identical to the one found in AdS, namely there is always a solution to Eq. (\ref{geneqM})
for any value of $\omega$.

We have seen that the results found in Minkowski space-time for the generation of marginally 
trapped surfaces are qualitatively the same as the AdS ones. 
For both $D=4$ and $D=5$ 
we have a critical value of the diluting parameter $\omega$ above which there are no solutions.
In $D=4$ we found that the critical behavior is of type I, i.e. the size of the trapped surface 
at criticality is finite, whereas for $D=5$ this critical size vanishes. For $D\geq 6$ there is 
no critical behavior in both cases. The main difference between 
the AdS and Minkowski cases lies in the value of the critical exponent for $D=5$, which is
$\gamma=1$ in AdS and $\gamma={1\over 2}$ in flat space-time. The reason for this
difference lies in the presence of the factor $1+2q_{\mathcal{C}}u$ in the integrand of 
Eq. (\ref{d5}). Because of it, the next-to-leading behavior of the integral on the left-hand side of
this equation is linear in $q_{\mathcal{C}}$ near the critical value of $\omega$. This is completely
different in Minkowski space-time, where the next-to-leading behavior of the integral in (\ref{d5m})
near $\omega^{*}$ is quadratic in $p_{\mathcal{C}}$. This difference is responsible for the change
in the value of the critical exponent.

\section{Concluding remarks}

In this paper we have studied the formation of marginally trapped surfaces in the head-on collision of two shock waves
as a function of the spread of energy in transverse space. For $D=4$ and $D=5$ we found respectively the existence of 
type I and type II criticality in both Minkowski and AdS space-time. This critical behavior is reminiscent of the 
one encountered in numerical simulations of gravitational collapse \cite{choptuikPRL,reviews}. 

In the lower dimensional case ($3<D<6$) we found that the type of trapped surfaces studied here is not formed when the 
incoming waves become too ``diluted" in energy. This, however, cannot be interpreted as implying that there is no apparent horizon, 
or eventually a black hole,
formed as the result of the head-on collision. To decide on this issue it would be necessary to solve the wave collision problem into
the interaction region $U>0$, $V>0$. However, it seems rather remarkable that when the trapped surface forms its size
follows a scaling rule with a well-defined critical exponent. The analysis we have presented here
shows that the values of these exponents
are quite independent of the particular form of the (rescaled) density function in transverse space.

It is interesting to notice that the Choptuik exponent for the formation of
trapped surfaces for shock waves in five dimensions agrees with the parton
saturation exponent computed for $\mathcal{N}=4$
super Yang Mills at strong coupling \cite{sym}. 
This agreement could be interpreted as some additional evidence on the
connection between black hole formation and parton saturation suggested in \cite{us}
in the strong coupling regime where the AdS/CFT correspondence is more reliable.

As a final remark, we would like to point out that the type of critical behavior described in this work takes place in a setup in which 
gravity is weak. This is an important difference with respect to critical behavior in black hole formation \cite{choptuikPRL,reviews}
where the curvature becomes large at the location of the forming black hole.

\section*{Acknowledgments} 

We would like to thank Steve Giddings for useful discussions. 
The work of C.G. has been partially supported by the Spanish
DGI contract FPA2003-02877 and the CAM grant HEPHACOS
P-ESP-00346. A.T. has been supported by Marie Curie and the Freydoon Mansouri foundations and
in part by NSF Grant PHY-0244821, and thanks the CERN Theory Group for
hospitality.
M.A.V.-M. acknowledges partial support from the Spanish
Government Grants PA2005-04823, FIS2006-05319, Basque Government Grant
IT-357-07 and Spanish Consolider-Ingenio
2010 Programme CPAN (CSD2007-00042).


\begin{thebibliography}{99}

\bibitem{choptuikPRL}
M.~W.~Choptuik,
{\it Universality and scaling in gravitational collapse of a massless scalar
field,}
Phys.\ Rev.\ Lett.\  {\bf 70} (1993) 9.


\bibitem{reviews}
C.~Gundlach and J.~M.~Mart\'{\i}n-Garc\'{\i}a,
{\it Critical phenomena in gravitational collapse,}
  Living Rev.\ Rel.\  {\bf 10} (2007) 5
{\tt   [arXiv:0711.4620 [gr-qc]].}



\bibitem{us}
L.~\'Alvarez-Gaum\'e, C.~G\'omez and M.~A.~V\'azquez-Mozo,
{\it Scaling Phenomena in Gravity from QCD},
Phys.\ Lett.\  {\bf B649} (2007) 478
{\tt  [arXiv:hep-th/0611312]}.
\\
L.~\'Alvarez-Gaum\'e, C.~G\'omez, A.~Sabio Vera, A.~Tavanfar and M.~A.~V\'azquez-Mozo,
{\it Critical gravitational collapse: towards a holographic understanding of the
  Regge region,}
  Nucl.\ Phys.\  {\bf B806} (2009) 327
{\tt  [arXiv:0804.1464 [hep-th]]}.


\bibitem{vw}
D.~Amati, M.~Ciafaloni and G.~Veneziano,
{\it Towards an S-matrix description of gravitational collapse,}
J. High Energy Phys. {\bf 02} (2008) 049
{\tt  [arXiv:0712.1209 [hep-th]].}
\\
G.~Veneziano and J.~Wosiek,
{\it Exploring an S-matrix for gravitational collapse,}
J. High Energy Phys. {\bf 09} (2008) 023
{\tt  [arXiv:0804.3321 [hep-th]].}
\\
G.~Veneziano and J.~Wosiek,
{\it Exploring an S-matrix for gravitational collapse II: a momentum space
analysis,}
J. High Energy Phys. {\bf 09} (2008) 024
{\tt  [arXiv:0805.2973 [hep-th]].}

\bibitem{acv}
D.~Amati, M.~Ciafaloni and G.~Veneziano,
{\it Superstring collisions at Planckian energies,}
  Phys.\ Lett.\  {\bf B197} (1987) 81.
\\
D.~Amati, M.~Ciafaloni and G.~Veneziano,
{\it Classical and quantum gravity effects from Planckian energy superstring
 collisions,}
Int.\ J.\ Mod.\ Phys.\  {\bf A3} (1988) 1615.
\\
D.~Amati, M.~Ciafaloni and G.~Veneziano,
{\it Can space-time be probed below the string size?}
  Phys.\ Lett.\  {\bf B216} (1989) 41.
\\
D.~Amati, M.~Ciafaloni and G.~Veneziano,
{\it Higher order gravitational deflection and soft bremsstrahlung in Planckian
energy superstring collisions,}
  Nucl.\ Phys.\  {\bf B347} (1990) 550.
\\
D.~Amati, M.~Ciafaloni and G.~Veneziano,
{\it Planckian scattering beyond the semiclassical approximation,}
  Phys.\ Lett.\  {\bf B289} (1992) 87.


\bibitem{gpy}
S.~S.~Gubser, S.~S.~Pufu and A.~Yarom,
{\it Entropy production in collisions of gravitational shock waves and of heavy
ions,}
  Phys.\ Rev.\  {\bf D78} (2008) 066014
{\tt  [arXiv:0805.1551 [hep-th]].}

\bibitem{GR}
D.~Grumiller and P.~Romatschke,
{\it On the collision of two shock waves in $AdS_5$,}
  J. High Energy Phys. {\bf 08} (2008) 027
{\tt  [arXiv:0803.3226 [hep-th]].}

\bibitem{penrose}
R.~Penrose, results presented at the Cambridge University Seminar, 1974 (unpublished).

\bibitem{dp}
P.~D.~D'Eath and P.~N.~Payne,
{\it Gravitational radiation in high speed black hole collisions. 1.
Perturbation treatment of the axisymmetric speed of light collision,}
  Phys.\ Rev.\  {\bf D46} (1992) 658.
  \\
   P.~D.~D'Eath and P.~N.~Payne,
{\it Gravitational radiation in high speed black hole collisions. 2. Reduction
to two independent variables and calculation of the second order news function,}
  Phys.\ Rev.\  {\bf D46} (1992) 675.
  \\
  P.~D.~D'Eath and P.~N.~Payne,
 {\it Gravitational radiation in high speed black hole collisions. 3. Results and
  conclusions,}
  Phys.\ Rev.\  {\bf D46} (1992) 694.

\bibitem{eardley_giddings}
D.~M.~Eardley and S.~B.~Giddings,
{\it Classical black hole production in high-energy collisions,}
  Phys.\ Rev.\  {\bf D66} (2002) 044011
{\tt  [arXiv:gr-qc/0201034].}

\bibitem{kv}
E.~Kohlprath and G.~Veneziano,
{\it Black holes from high-energy beam-beam collisions,}
  J. High Energy Phys {\bf 06} (2002) 057
{\tt  [arXiv:gr-qc/0203093].}

\bibitem{poisson}
E.~Poisson, {\it A Relativist's Toolkit: The Mathematics of Black-Hole Mechanics}, 
Cambridge 2004.

\bibitem{sym}
Y.~Hatta, E.~Iancu and A.~H.~Mueller,
{\it Deep inelastic scattering at strong coupling from gauge/string duality:
the saturation line,}
  J. High Energy Phys. {\bf 01} (2008) 026
 {\tt  [arXiv:0710.2148 [hep-th]].}
  \\
  E.~Levin, J.~Miller, B.~Z.~Kopeliovich and I.~Schmidt,
{\it Glauber-Gribov approach for DIS on nuclei in $\mathcal{N}=4$ SYM,}
  {\tt arXiv:0811.3586 [hep-ph].}






\end{thebibliography}
\end{document}